# ECONOMIC GROWTH MODEL
# WITH CONSTANT PACE AND DYNAMIC MEMORY

**Valentina V. Tarasova**,

Higher School of Business, Lomonosov Moscow State University,
Moscow 119991, Russia; E-mail: v.v.tarasova@mail.ru;

**Vasily E. Tarasov**,

Skobeltsyn Institute of Nuclear Physics, Lomonosov Moscow State University,
Moscow 119991, Russia; E-mail: tarasov@theory.sinp.msu.ru

**Abstract:** *The article discusses a generalization of model of economic growth with constant pace, which takes into account the effects of dynamic memory. Memory means that endogenous or exogenous variable at a given time depends not only on their value at that time, but also on their values at previous times. To describe the dynamic memory we use derivatives of non-integer orders. We obtain the solutions of fractional differential equations with derivatives of non-integral order, which describe the dynamics of the output caused by the changes of the net investments and effects of power-law fading memory.*

**Introduction**

In the continuous time approach, the economic growth models are described by using the tool of differential equations with derivatives of integer orders [1, 2, 3]. In mathematics, derivatives of non-integer order are also well known [4, 5]. This tool allows us to describe processes with power-law memory (for example, see [6]). In this paper, we will consider a simplest economic model of growth with dynamic memory. We propose a generalization of economic growth model with constant pace. We first describe the standard model, which does not take into account the effects of time delay and memory. Let Y(t) be a function that describes the volume of production (the output), which was produced and sold at time t. We will use the assumption of unsaturation of the consumer market, i.e. we will assume that all made production is sold. In the simplest case, we also can assume that the sales volume is not so high as to significantly affect the price P. This allows us to consider a fixed price (P(t)=P).

It is known that an increase of the production volume Y(t) is caused by the net investments I(t), which is investments aimed at expansion of production. The amount of net investment equal to the difference between the total investment and amortization (depreciation) costs. To increase output it is necessary that the net investment I(t) is greater than zero (I(t)>0). In the case I(t)=0, the investments only cover the cost of amortization and the output level remains unchanged. In the case I(t)<0, we have a reduction of fixed assets and, as a consequence, a decrease of output.

The growth model with constant pace is assumed that the marginal output (dY(t)/dt) is directly proportional to the net investment I(t). Mathematically, it is written by the differential equation

$$\frac{dY(t)}{dt} = L \cdot I(t), \qquad (1)$$

where L is the rate of acceleration [3]. Assuming that the amount of investment I(t) is a fixed part of income Q(t)=P·Y (t), we obtain

$$I(t) = m \cdot P \cdot Y(t), \qquad (2)$$

where m is the norm of the net investment (0<m<1), which describes a part of income that is spent on the net investment. Substituting expression (2) into equation (1), we obtain

$$\frac{dY(t)}{dt} = \lambda \cdot Y(t), \qquad (3)$$

where λ=m·P·L. Differential equation (3) has the solution

$$Y(t) = Y(0) \cdot \exp(\lambda \cdot t). \qquad (4)$$

Differential equation (3) describes the increase of output without restriction of growth [3, p. 81]. This equation is equation of growth with a constant pace.

Equations (1) and (3) contain only the first-order derivative of Y(t) with respect to time. It is known that the derivatives of integer orders are determined by the properties of differentiable functions of time only in infinitely small neighborhood of the considered point of time. As a result, this economic model, which is described by equation (3), assumes an instant changes of marginal output, when the net investment changes. This means that the effects of dynamics memory and lag are neglected. The dynamic memory means a dependence of output at the present time on the investment changes in the past. In other words, equation (3) does not take into account the effects of memory. In economic models, we can consider the concept of dynamics memory by analogy with this concept in physics [6, p. 394-395]. The term "memory" means that the process state at a given time t=T depends on the process state in the past (t<T). In economic processes, a presence of memory means that there is endogenous or exogenous variable, which depends not only on its values at present time, but also on its values at previous time points [7, 8]. A memory effect is related with the fact that the same change of the exogenous variable can leads to the different change of the corresponding endogenous variable. This leads us to the multivalued dependencies of these variables [7, 8]. The multivalued dependencies are caused by the fact that the economic agents remember previous changes of these variables, and therefore can react differently. As a result, identical changes of the exogenous variable may lead to the different dynamics of endogenous variable. To describe power-law memory we can use the theory of derivatives and integrals of non-integer order [4, 5]. An economic interpretation of the fractional derivatives has been suggested in [14, 15]. To take into account the effects of power-law memory, the concept of marginal values of non-integer order [9, 10, 11] and the concept of the accelerator with memory [12, 13] have been proposed. The economic dynamic with power-law memory can be described by using the fractional differential equations [14, 15, 16, 17, 18, 19, 20] with derivatives of non-integer orders.

**Case of dynamic memory with power-law fading**

In mathematics different types of fractional-order derivatives are known [4]. In order to take into account a power-law dynamic memory, we propose to use the left-sided Caputo fractional derivative of order α>0 with respect to time. One of the important properties of the Caputo fractional derivatives is that the action of these derivatives on a constant function gives zero. Using only the left-sided fractional-order derivative, we take into account the history of changes of endogenous or exogenous variable in the past, that is for t<T. The right-sided Caputo derivatives are defined by integration over t>T. The left-sided Caputo fractional derivative is defined by the equation

$$(D_{0+}^\alpha Y)(t) := \frac{1}{\Gamma(n-\alpha)} \int_0^t \frac{Y^{(n)}(\tau)d\tau}{(t-\tau)^{\alpha-n+1}}, \tag{5}$$

where $Y^{(n)}(\tau)$ is the derivative of integer order n=[α]+1 of the function $Y(\tau)$ with respect to τ such that 0<τ<t. Here function $Y(\tau)$ must have the derivatives of integer orders up to the (n-1)th order, which are absolutely continuous functions on the interval [0,t]. In order to have the correct dimensions of economic quantities we will use the dimensionless time variable t.

Using the concept of the marginal values of non-integer orders, which is suggested in [9, 10, 11], and the concept of the accelerator with memory, which is proposed in [12], we get a generalization of equation (3) in the form

$$(D_{0+}^\alpha Y)(t) = \lambda \cdot Y(t). \tag{6}$$

Fractional differential equation (6) takes into account one-parametric memory with power-law fading. Let us consider the Cauchy problem for fractional differential equation (6) of order α > 0 with the initial conditions

$$Y^{(k)}(0) = Y_k, \tag{7}$$

where n–1<α<n, λ is a real number, and k=1,…,n–1.

For this Cauchy problem, we can give the conditions for a unique solution Y(t) in the space $C_\gamma^{\alpha,n-1}[0,T]$, where 0≤t≤T, 0≤γ<1 and γ≤α. This function space is defined by

$$C_\gamma^{\alpha,n-1}[0,T] = \{Y(t) \in C^n[0,T]: (D_{0+}^\alpha Y)(t) \in C_\gamma[0,T]\}, \tag{8}$$

where $C_\gamma[0,T]$ is the weighted space of functions Y(t) given on [0,T ], such that $t^\gamma \cdot Y(t) \in C[0,T]$. The space $C^n[0,T]$ is the space of functions Y(t), which are continuously differentiable on [0,T] up to order n. The space C[0,T] is the space of functions Y(t), which are continuous on [0,T].

Using Theorem 4.3 of [4, p. 231], the Cauchy problem involving homogeneous fractional differential equation (6) and initial conditions (7) has a unique solution $Y(t) \in C_\gamma^{\alpha,n-1}[0,T]$ in the form

$$Y(t) = \sum_{k=0}^{n-1} Y_k \cdot t^k \cdot E_{\alpha,k+1}[\lambda \cdot t^\alpha], \tag{9}$$

where $E_{\alpha,\beta}[z]$ is the two-parameter Mittag-Leffler function [4, p. 42], which is defined by the equation

$$E_{\alpha,\beta}[z] := \sum_{k=0}^\infty \frac{z^k}{\Gamma(\alpha k+\beta)}. \tag{10}$$

The Mittag-Leffler function $E_{\alpha,\beta}[z]$ is a generalization of the exponential function $e^z$, since $E_{1,1}[z] = e^z$. Solution (9) describes the economic growth model with constant pace and power-law fading memory.

For 0<α<1, the solution of equation (6) has the form

$$Y(t) = Y(0) \cdot E_{\alpha,1}[\lambda \cdot t^\alpha]. \tag{11}$$

For α=1, equation (11) gives solution (4), which describes the economic growth model without memory.

**Case of power- law price and memory**

Let us consider the case, when the price P=P(t) is changed according to the power law

$$P(t) = p \cdot t^\beta, \tag{12}$$

where β≥0 and p>0. In this case, we have the fractional differential equation

$$(D_{0+}^\alpha Y)(t) = \lambda \cdot t^\beta \cdot Y(t), \tag{13}$$

where the coefficient $\lambda$ is defined by the equation $\lambda = m \cdot p \cdot L$.

Using Theorem 4.4 of [4, p. 233], the Cauchy problem involving fractional differential equation (13) and initial conditions (7) has a unique solution $Y(t) \in C_\gamma^{\alpha,n-1}[0,T]$ in the form

$$Y(t) = \sum_{k=0}^{n-1} \frac{1}{k!} Y_k \cdot t^k \cdot E_{\alpha, 1+\beta/\alpha, (\beta+k)/\alpha}[\lambda \cdot t^{\alpha+\beta}], \tag{14}$$

where $E_{\alpha,b,c}[z]$ is the generalized Mittag-Leffler function [4, p. 48]. This function is defined by the equation

$$E_{\alpha,b,c}(z) := \sum_{k=0}^\infty a_k(\alpha, b, c) \cdot z^k, \tag{15}$$

where $a_0(\alpha, b, c) = 1$ and

$$a_k(\alpha, b, c) = \prod_{j=0}^{k-1} \frac{\Gamma(\alpha(bk+c)+1)}{\Gamma(\alpha(bk+c+1)+1)} \tag{16}$$

for integer $k \geq 1$.

For $\beta = 0$, we have $E_{\alpha,1,k/\alpha}[\lambda \cdot t^\alpha] = k! \cdot E_{\alpha,k+1}[\lambda \cdot t^\alpha]$. Therefore equation (14) with $\beta = 0$ gives (9).

For $0 < \alpha < 1$, the solution of equation (14) has the form

$$Y(t) = Y(0) \cdot E_{\alpha, 1+\beta/\alpha, \beta/\alpha}[\lambda \cdot t^{\alpha+\beta}], \tag{17}$$

where we get (11) for the case $\beta = 0$.

**Case of two-parameter power-law memory**

Let us consider model with two-parameter power-law memory. The differential equation of the growth model with this memory has the form

$$(D_{0+}^\alpha Y)(t) - \mu \cdot (D_{0+}^\beta Y)(t) = \lambda \cdot Y(t), \tag{18}$$

where $\alpha > \beta > 0$, $n-1 < \alpha \leq n$, $m-1 < \beta \leq m$, $m \leq n$, $0 \leq t \leq T$, and $\mu, \lambda$ are real number. The solution of (18) is represented in terms of the generalized Wright function (the Fox-Wright function), $\Psi_{1,1}\begin{bmatrix}(a,\alpha)\\(b,\beta)\end{bmatrix}z$, which is defined by the equation

$$\Psi_{1,1}\begin{bmatrix}(a,\alpha)\\(b,\beta)\end{bmatrix}z := \sum_{k=0}^\infty \frac{\Gamma(\alpha \cdot k + a)}{\Gamma(\beta \cdot k + b)} \cdot \frac{z^k}{k!}. \tag{19}$$

Using Theorem 5.13 of [4, p.314], the solution of equation (18) has the form

$$Y(t) = \sum_{j=0}^{n-1} a_j Y_j(t), \tag{20}$$

where $Y_j(t)$, $j = 0, \ldots, n-1$ are defined by the following equations

$$Y_j(t) = \sum_{k=0}^\infty \frac{\lambda^k \cdot t^{k\alpha+j}}{\Gamma(k+1)} \Psi_{1,1}\begin{bmatrix}(n+1,1)\\(\alpha k+j+1, \alpha-\beta)\end{bmatrix}\mu \cdot t^{\alpha-\beta}\Big] -$$

$$\mu \cdot \sum_{k=0}^\infty \frac{\lambda^k \cdot t^{k\alpha+j+\alpha-\beta}}{\Gamma(k+1)} \Psi_{1,1}\begin{bmatrix}(n+1,1)\\(\alpha k+j+1+\alpha-\beta, \alpha-\beta)\end{bmatrix}\mu \cdot t^{\alpha-\beta}\Big] \tag{21}$$

for $j = 0, \ldots, m-1$, and

$$Y_j(t) = \sum_{k=0}^\infty \frac{\lambda^k \cdot t^{k\alpha+j}}{\Gamma(k+1)} \Psi_{1,1}\begin{bmatrix}(n+1,1)\\(\alpha k+j+1, \alpha-\beta)\end{bmatrix}\mu \cdot t^{\alpha-\beta}\Big] \tag{22}$$

for $j = m, \ldots, n-1$.

For $0 < \beta < \alpha \leq 1$, the solution of equation (18) is written in the form

$$Y(t) = \sum_{k=0}^\infty \frac{\lambda^k \cdot t^{k\alpha}}{\Gamma(k+1)} \Psi_{1,1}\begin{bmatrix}(n+1,1)\\(\alpha k+1, \alpha-\beta)\end{bmatrix}\mu \cdot t^{\alpha-\beta}\Big] -$$

$$\mu \cdot \sum_{k=0}^\infty \frac{\lambda^k \cdot t^{k\alpha+\alpha-\beta}}{\Gamma(k+1)} \Psi_{1,1}\begin{bmatrix}(n+1,1)\\(\alpha k+1+\alpha-\beta, \alpha-\beta)\end{bmatrix}\mu \cdot t^{\alpha-\beta}\Big]. \tag{23}$$

For $1 < \beta < \alpha \leq 2$, the solution of equation (18) has the form

$$Y(t) = a_0 Y_0(t) + a_1 Y_1(t), \tag{24}$$

where $Y_0(t)$ is defined by (23), and $Y_1(t)$ is defined by the equation

$$Y_1(t) = \sum_{k=0}^{\infty} \frac{\lambda^k \cdot t^{k\alpha+1}}{\Gamma(k+1)} \Psi_{1,1}\left[{(n+1,1) \atop (\alpha k+2, \alpha-\beta)} | \mu \cdot t^{\alpha-\beta}\right] -$$
$$\mu \cdot \sum_{k=0}^{\infty} \frac{\lambda^k \cdot t^{k\alpha+1+\alpha-\beta}}{\Gamma(k+1)} \Psi_{1,1}\left[{(n+1,1) \atop (\alpha k+2+\alpha-\beta, \alpha-\beta)} | \mu \cdot t^{\alpha-\beta}\right]. \quad (25)$$

For $0<\beta<1<\alpha\leq 2$, the solution of equation (18) is represented by equation (24) with $Y_0(t)$ in the form (23), and $Y_1(t)$ that is defined by the equation

$$Y_1(t) = \sum_{k=0}^{\infty} \frac{\lambda^k \cdot t^{k\alpha+1}}{\Gamma(k+1)} \Psi_{1,1}\left[{(n+1,1) \atop (\alpha k+2, \alpha-\beta)} | \mu \cdot t^{\alpha-\beta}\right]. \quad (26)$$

For the case of the multi-parametric power-law memory, we can use Theorem 5.14 of [4, p. 319-320]. Two-parametric and multi-parametric memory allows us to take into account the power-law fading of memory for different types of economic agents.

**Dynamics of price growth and fixed assets with memory**

Some economic processes can be described by the analogous equations. For example, such processes are the price growth at a constant pace of inflation and dynamics of fixed assets.

Let us consider the dynamics of price growth at a constant pace of inflation. We will assume that the price at time t is equal to P(t). The inflation pace is assumed to be equal to the constant R. Then, the price growth with power-law memory at constant pace of inflation can be described by the fractional differential equation

$$(D_{0+}^\alpha P)(t) = R \cdot P(t), \quad (27)$$

where $D_{0+}^\alpha$ is the Caputo derivative (5). For $\alpha=1$, equation (27) takes the form

$$\frac{dP(t)}{dt} = R \cdot P(t). \quad (28)$$

Fractional differential equation (27) has the solution

$$P(t) = \sum_{k=0}^{n-1} P_k \cdot t^k \cdot E_{\alpha,k+1}[R \cdot t^\alpha], \quad (29)$$

where $E_{\alpha,\beta}[z]$ is the two-parameter Mittag-Leffler function (10). Solution (29) describes the dynamics of price growth with power-law fading memory. For $\alpha=1$, expression (29) takes the form

$$P(t) = P(0) \cdot \exp(R \cdot t), \quad (30)$$

which is the solution of equation (28), which describes the price growth at a constant pace [3, p. 81] without memory effects.

As a second example we consider the dynamics of fixed assets, where we take into account the memory effects. Let B be a coefficient of disposal of fixed assets. We assume that the investment is constant, which is equal to A monetary units. We can describe the dynamics of fixed assets, if the rate of change of the fixed assets is equal to the difference between investments and disposal of fixed assets. Let us denote the fixed assets at time t≥0 by K(t). The dynamics of the fixed assets with power-law memory can be described by the fractional differential equation

$$(D_{0+}^\alpha K)(t) = A - B \cdot K(t), \quad (31)$$

where $D_{0+}^\alpha$ is the Caputo derivative (5). For $\alpha=1$, equation (31) takes the form

$$\frac{dK(t)}{dt} = A - B \cdot K(t). \quad (32)$$

Equation (32) describes the dynamics of fixed assets [3, p. 82] without memory.
The solution of equation (31) has [4, p. 323] the form

$$K(t) = A \cdot \int_0^t (t-\tau)^{\alpha-1} \cdot E_{\alpha,\alpha}[-B \cdot (t-\tau)^\alpha] d\tau +$$
$$\sum_{k=0}^{n-1} K^{(k)}(0) \cdot t^k \cdot E_{\alpha,k+1}[-B \cdot t^\alpha], \quad (33)$$

where n-1<α≤n, $E_{\alpha,\beta}[z]$ is the two-parameter Mittag-Leffler function (10). The calculation of the integral in equation (33) by using the change of variable ξ = t-τ, the definition (10) of the Mittag-Leffler function and term by term integration, gives solution (33) in the form

$$K(t) = \frac{A}{B} \cdot \left(1 - E_{\alpha,1}[-B \cdot t^\alpha]\right) + \sum_{k=0}^{n-1} K^{(k)}(0) \cdot t^k \cdot E_{\alpha,k+1}[-B \cdot t^\alpha], \qquad (34)$$

where n-1<α≤n, and $K^{(k)}(0)$ are the values of the derivatives of the function K(t) at t=0. Solution (34) describes the dynamics of fixed assets with power-law fading memory.

For 0<α≤1 (n=1) solution (34) has the form

$$K(t) = \frac{A}{B} \cdot \left(1 - E_{\alpha,1}[-B \cdot t^\alpha]\right) + K(0) \cdot E_{\alpha,k+1}[-B \cdot t^\alpha]. \qquad (35)$$

Using $E_{1,1}[z] = e^z$, solution (35) with α=1 takes the form

$$K(t) = \frac{A}{B}(1 - \exp(-B \cdot t)) + K(0) \cdot \exp(-B \cdot t), \qquad (36)$$

which describes the dynamics (32) of fixed assets without memory effects.

**Conclusion**

In general, in economic models we should take into account the memory effects that are based on the fact that economic agents remember the history of changes of exogenous and endogenous variables that characterize the economic process. The proposed economic growth model with constant pace and power-law memory has shown that the memory effects can play an important role in economic phenomena and processes.

**References**
1. Allen R.G.D. Mathematical Economics. Second edition. London: Macmillan, 1960. 812 p.
2. Allen R.G.D. Macro-Economic Theory. A Mathematical Treatment. London: Macmillan, 1968. 420 p.
3. Volgina O.A., Golodnaya N.Y., Odiyako N.N., Schumann G.I. Mathematical Modeling of Economic Processes and Systems. 3rd ed. Moscow: Cronus, 2014. 200 p. [in Russian]
4. Kilbas A.A., Srivastava H.M., Trujillo J.J. Theory and Applications of Fractional Differential Equations. Amsterdam: Elsevier, 2006. 540 p.
5. Diethelm K. The Analysis of Fractional Differential Equations: An Application-Oriented Exposition Using Differential Operators of Caputo Type. Berlin: Springer-Verlag, 2010. 247 p.
6. Tarasov V.E. Fractional Dynamics: Applications of Fractional Calculus to Dynamics of Particles, Fields and Media. New York: Springer, 2010. 505p.
7. Tarasova V.V., Tarasov V.E. Criteria of hereditarity of economic process and the memory effect // Molodoj Uchenyj [Young Scientist], 2016. No. 14 (118). P. 396-399. [in Russian].
8. Tarasov V.E., Tarasova V.V. Long and short memory in economics: fractional-order difference and differentiation // IRA-International Journal of Management and Social Sciences, 2016. Vol. 5. No. 2. P. 327–334. DOI: 10.21013/jmss.v5.n2.p10 (arXiv:1612.07913)
9. Tarasova V.V., Tarasov V.E. Marginal utility for economic processes with memory // Almanah Sovremennoj Nauki i Obrazovaniya [Almanac of Modern Science and Education], 2016. No. 7 (109). P. 108-113. [in Russian].
10. Tarasova V.V., Tarasov V.E. Marginal values of non-integer order in the economic analysis // Azimut Nauchnih Issledovanii: Ekonomika i Upravlenie [Azimuth Research: Economics and Management], 2016. No. 3 (16). P. 197-201. [in Russian].

11. Tarasova V.V., Tarasov V.E. Economic indicator that generalizes average and marginal values // Ekonomika i Predprinimatelstvo [Journal of Economy and Entrepreneurship], 2016. No. 11–1 (76–1). P. 817–823. [in Russian].
12. Tarasova V.V., Tarasov V.E. A generalization of the concepts of the accelerator and multiplier to take into account of memory effects in macroeconomics // Ekonomika i Predprinimatelstvo [Journal of Economy and Entrepreneurship], 2016. No. 10–3 (75-3). P. 1121–1129. [in Russian].
13. Tarasova V.V., Tarasov V.E. Economic accelerator with memory: discrete time approach // Problems of Modern Science and Education, 2016. No. 36 (78). P. 37–42. DOI: 10.20861/2304-2338-2016-78-002 (arXiv:1612.07913)
14. Tarasova V.V., Tarasov V.E. Elasticity for economic processes with memory: Fractional differential calculus approach // Fractional Differential Calculus, 2016. Vol. 6. No. 2. P. 219-232. DOI: 10.7153/fdc-06-14
15. Tarasova V.V., Tarasov V.E. Logistic map with memory from economic model // Chaos, Solitons and Fractals, 2017. Vol. 95. P. 84-91. DOI: 10.1016/j.chaos.2016.12.012
16. Tarasova V.V., Tarasov V.E. Fractional dynamics of natural growth and memory effect in economics // European Research, 2016. No. 12 (23). P. 30-37. DOI: 10.20861/2410-2873-2016-23-004 (arXiv:1612.09060)
17. Tarasova V.V., Tarasov V.E. Hereditarity generalization of Harrod-Domar model and memory effects // Ekonomika i Predprinimatelstvo [Journal of Economy and Entrepreneurship], 2016. No. 10-2 (75-2). P. 72-78.
18. Tarasova V.V., Tarasov V.E Memory effects in hereditary model Harrod-Domar // Problemy Sovremennoj Nauki i Obrazovanija [Problems of Modern Science and Education], 2016. No. 32 (74). P. 38-44. DOI: 10.20861/2304-2338-2016-74-002
19. Tarasova V.V., Tarasov V.E Keynesian model of economic growth with memory // Ekonomika i Upravlenie: Problemy Resheniya [Economy and Management: Problems and Solutions], 2016. No. 10-2 (58). P. 21-29.
20. Tarasova V.V., Tarasov V.E Memory effects in hereditary Keynesian model // Problemy Sovremennoj Nauki i Obrazovanija [Problems of Modern Science and Education], 2016. No. 38 (80). P. 56-61. DOI: 10.20861/2304-2338-2016-80-001